\documentclass[12pt,epsf,amssymb,ulem,qsymbols]{article}
\usepackage{tabularx}
\usepackage{array}
\usepackage{graphics}
\usepackage{graphicx}
\usepackage{psfrag}
\usepackage{epsfig}
\usepackage{amsmath}
\usepackage{amssymb}
\usepackage{ulem}
\usepackage{setspace}
\usepackage{rotating}
\usepackage{colortbl}
\usepackage{tabularx}
\usepackage{longtable}
\usepackage{multirow}
\makeatletter

\usepackage{verbatim}

\setlongtables

\newcommand{\msbar}{{\overline{\rm MS}}}

\newcommand{\bea}{\begin{eqnarray}}
\newcommand{\eea}{\end{eqnarray}}
\newcommand{\beq}{\begin{equation}}
\newcommand{\eeq}{\end{equation}}
\newcommand{\ec}{\end{center}}
\newcommand{\bc}{\begin{center}}
\newcommand{\gev}{{\rm GeV}}
\newcommand{\mev}{{\rm MeV}}

\newcommand{\pdir}{p\kern -5.2pt\raise 0.2ex\hbox {/}}

\newcommand{\vdir}{v\kern -5.75pt\raise 0.15ex\hbox {/}}
\newcommand{\kdir}{k\kern -5.75pt\raise 0.15ex\hbox {/}}
\newcommand{\epsdir}{\epsilon\kern -5.0pt\raise 0.15ex\hbox {/}}
\newcommand{\bvdir}{\bar{v}\kern -5.75pt\raise 0.15ex\hbox {/}}
\newcommand{\Ddir}{D\kern -7.75pt\raise 0.20ex\hbox {/}}
\newcommand{\Adir}{A\kern -7.75pt\raise 0.20ex\hbox {/}}
\newcommand{\ldir}{l\kern -5.0pt\raise 0.2ex\hbox{/}}
\newcommand{\varepsdir}{\varepsilon\kern -5.5pt\raise 0.15ex\hbox{/}}

\newcommand{\nn}{\nonumber}
\makeatother

\begin{document}
\thispagestyle{empty} 
\begin{flushright}
\begin{tabular}{l}
{\tt LPT 12-52}\\
\end{tabular}
\end{flushright}
\begin{center}
\vskip 2.8cm\par
{\par\centering \textbf{\LARGE  
\Large \bf Lattice QCD study of the radiative decays }}\\
\vskip .4cm\par
{\par\centering \textbf{\LARGE  
\Large \bf $J/\psi\to \eta_c\gamma$  and $h_c\to \eta_c\gamma$  }}\\
\vskip 1.05cm\par
{\scalebox{.8}{\par\centering \large  
\sc Damir Be\v{c}irevi\'c$^a$ and Francesco Sanfilippo$^{a,b}$}
{\par\centering \vskip 0.65 cm\par}
{\sl 
$^a$~Laboratoire de Physique Th\'eorique (B\^at.~210)~\footnote{Laboratoire de Physique Th\'eorique est une unit\'e mixte de recherche du CNRS, UMR 8627.}\\
Universit\'e Paris Sud, F-91405 Orsay-Cedex, France.}\\
{\par\centering \vskip 0.25 cm\par}
{\sl 
$^b$~INFN, Sezione di Roma, \\
Piazzale Aldo Moro 5, I-00185 Roma, Italy.}\\ 
{\vskip 1.65cm\par}}
\end{center}

\vskip 0.55cm
\begin{abstract}
We present  the results of our lattice QCD study of the hadronic matrix elements relevant to the physical 
radiative $J/\psi\to \eta_c\gamma$  and $h_c\to \eta_c\gamma$ decays. We used the twisted mass QCD action 
with $N_{\rm f}=2$ light dynamical quarks and from the computations made at four lattice spacings we were able to 
take the continuum limit. Besides the form factors parameterizing the above decays we also computed: (i) the hyperfine splitting 
and obtained $\Delta = 112\pm 4$~MeV,  (ii) the annihilation constant $f_{J/\psi}$ which agrees with the one 
inferred from the measured $\Gamma(J/\psi \to e^+e^- )$.
\end{abstract}
\newpage
\setcounter{page}{1}
\setcounter{footnote}{0}
\setcounter{equation}{0}
\noindent

\renewcommand{\thefootnote}{\arabic{footnote}}

\setcounter{footnote}{0}
\section{\label{sec-0}Introduction}
After the observation of  $\eta_b$ at BaBar~\cite{eta-exp-babar} the radiative decays of heavy quarkonia received a significant attention in the literature. As $\Upsilon(1S) \to \eta_b\gamma$ is not yet experimentally accessible due to the smallness of phase space, the experimenters turned to studying  $\Upsilon(2S) \to \eta_b\gamma$ and  $\Upsilon(3S) \to \eta_b\gamma$ modes from which they then extracted $m_{\eta_b}$.  A potential problem in that value is the insufficient control of theoretical uncertainties in the transition matrix elements to guarantee an accuracy at a percent level, especially when the radial excitations are involved. The corresponding transition matrix elements have been computed by using quark models~\cite{QMsummary}.~\footnote{For a complete reviews with extensive lists of references please see ref.~\cite{QMsummary,onia-review}.}  Indeed the resulting value $m_{\eta_b}= 9390.9\pm 2.1$~MeV extracted from the BaBar experiment~\cite{eta-exp-babar} was consistent with the value obtained from the similar measurements made at CLEO~\cite{eta-exp-cleo}, leading to the following value of the hyperfine splitting~\cite{PDG}, 
\bea\label{eq:hfs}
\Delta_b^{\rm exp.}= m_{\Upsilon(1S)}- m_{\eta_b} =69.3\pm 2.8\ \mev\,,
\eea
that turned out to be much larger than the values predicted by methods based on perturbative QCD, namely $\Delta_b=44\pm 11$~MeV~\cite{Recksiegel:2003fm}, $39\pm 14$~MeV~\cite{Kniehl:2003ap}. The lattice QCD results are inconclusive on this issue so far, although they seem to point towards the values larger than those obtained in refs.~\cite{Recksiegel:2003fm,Kniehl:2003ap}. For example, by using the simulations with $N_{\rm f}=2+1$ staggered quarks and the Fermilab treatment of the heavy quarks on the lattice, the value $\Delta_b^{\rm latt}=54\pm 12$~MeV was obtained in ref.~\cite{0912}, while the non-relativistic QCD (NRQCD) treatment of the heavy quarks lead to $\Delta_b^{\rm latt}=70\pm 9$~MeV~\cite{dawdal}. Simulations of QCD with $N_{\rm f}=2+1$ light flavors of the domain wall light quark flavors and by using NRQCD for the heavy, resulted in  $\Delta_b^{\rm latt}=60\pm 8$~MeV~\cite{meinel}.

The experimenters at Belle avoided using radial excitations and from a large sample of $h_b(1P)$~\cite{Adachi:2011ji} they were able to  measure the $h_b(1P)\to \eta_b \gamma$ decay rate, which resulted in a somewhat larger value $m_{\eta_b}= 9401.0\pm 1.9^{+1.4}_{-2.4}$~MeV~\cite{belle-hb} (i.e. smaller $\Delta_b^{\rm exp.} =59.6\pm 2.7$~MeV), but still lighter than expected from the models~\cite{QMsummary} or the analytic calculations of refs~\cite{Recksiegel:2003fm,Kniehl:2003ap}. 

Proponents of the extensions of the Standard Model involving more than one Higgs doublet speculated that the experimentally established pseudoscalar state $\eta_b$ might be actually a mixture of the true $\eta_b$ and the light parity-odd Higgs boson $A^0$~\cite{A0}.~\footnote{An abridged discussion on this issue with a more complete list of references can be found in ref.~\cite{domingo}. } This would solve the puzzle of too large a hyperfine splitting and would give more support to a plausible solution that  $m_{A^0}\sim 9$~GeV. However,  to give these speculations more support it is essential to check whether or not the hadronic matrix element used to extract $m_{\eta_b}$ from the mentioned experiments coincides with the results obtained by using the methods based on QCD from first principles. For example, by using NRQCD on the lattice, the authors of ref.~\cite{lattice-radiative-3} obtained much larger values for the transition matrix elements than those inferred from the measured $
\Upsilon(nS)\to \eta_b\gamma$ ($n=2,3$).  Since the direct QCD simulations of the $b\bar b$- systems are difficult because the lattice spacings are still too large to resolve the $b$-quark mass,  we decided to explore the similar physics processes in the charmed systems and study, $J/\psi \to \eta_c\gamma$ and $h_c(1P)\to \eta_c\gamma$. The established methodology of this paper will then be used for our future attempt to compute the amplitude for  $h_b(1P)\to \eta_b \gamma$ decay on the lattice. Besides methodological issues these decays are physically interesting on their own. One is the so called magnetic dipole (M1) and the other electric dipole decay (E1). Quark models fail to reproduce the measured $\Gamma(J/\psi \to \eta_c\gamma)$ and,  instead, obtain a significantly larger value~\cite{QMsummary}.  On the other hand, $h_c(1P)$ has been discerned from the experimental background only recently and its dominant decay is indeed $h_c(1P)\to \eta_c\gamma$, the branching fraction of which has been measured accurately. 

The first extensive study of the radiative decays of charmonia on the lattice has been reported in ref.~\cite{lattice-radiative-1} where the authors computed relevant matrix elements for a number of decay channels in the quenched approximation of QCD and with one lattice spacing. That computation has been extended to the case of $N_{\rm f}= 2$ dynamical light quark flavors at single lattice spacing~\cite{lattice-radiative-2}. In this paper we will focus on $J/\psi \to \eta_c\gamma$ and $h_c(1P)\to \eta_c\gamma$, for which we compute the desired form factors at four lattice spacings and then extrapolate them to the continuum limit. Those results may be used for a cleaner exclusion of the possibility of having very light $m_{A^0}<2 m_\tau$ (see e.g.~\cite{A0}).

\section{\label{sec-2}Hadronic Matrix Elements }

The transition matrix element responsible for the $J/\psi\to \eta_c\gamma^\ast$ decay reads, 
\bea\label{def-vectorFF}
\langle \eta_c(k) \vert  J^{\rm em}_\mu \vert J/\psi(p,\epsilon_\lambda) \rangle = 
e {\cal Q}_c\ \varepsilon_{\mu\nu\alpha\beta}\  \epsilon_\lambda^{\ast \nu} p^\alpha k^\beta \ \frac{2\ V(q^2)}{ m_{J/\psi}+m_{\eta_c} }\,,
\eea
where $J^{\rm em}_\mu ={\cal Q}_c \bar c \gamma_\mu c$ is the relevant piece of the electromagnetic current, with  ${\cal Q}_c=2/3$ in units of $e=\sqrt{4\pi \alpha_{\rm em}}$. Information regarding the non-perturbative QCD dynamics is encoded in the form factor $V(q^2)$ and represents the most challenging part on the theory side.  For the physical process, i.e. with the photon on-shell $q^2=0$, the decay rate is given by~\cite{lattice-radiative-1}
\bea\label{widthPSI}
\Gamma(J/\psi\to \eta_c\gamma)& = & {64\over 27} \   {    \alpha_{\rm em} \  \vert  \  \vec   q \  \vert^3 \over (m_{J/\psi}+m_{\eta_c} )^2}    \left| V(0)\right|^2 \cr
& = & {8\over 27}\ \alpha_{\rm em}\   (m_{J/\psi}+m_{\eta_c} )\  \left({\Delta \over m_{J/\psi}}\right)^3 \left| V(0)\right|^2\,,
\eea
where $\Delta$ stands for the hyperfine splitting $\Delta=m_{J/\psi}-m_{\eta_c} $. 
When both the initial and final hadrons are at rest the matrix element~(\ref{def-vectorFF}) is zero by definition. The smallest momentum that can be given to a hadron on the lattice with periodic boundary conditions is $2\pi/L$, which is very large for the lattices that we work with today and would make $q^2 <0$, far from $q^2=0$. As a result we would have to  work at several negative $q^2$'s, then model the $q^2$ shape of the form factor as to extrapolate to the physical point, $q^2=0$. That methodology has been adopted in refs.~\cite{lattice-radiative-1,lattice-radiative-2}. In this work, instead, we will use the so called twisted boundary conditions~\cite{twbc} which allow us to work directly at $q^2=0$. This is achieved by tuning the twisting angle $\theta_0$ via the three momentum given to the pseudoscalar meson that fulfills the condition,
\bea\label{eq:angle1}
|\vec q| = {m_{J/\psi}^2-m_{\eta_c}^2 \over 2 m_{J/\psi}}\; \Rightarrow  \; \theta_0 = {L\over \sqrt{3}} {m_{J/\psi}^2-m_{\eta_c}^2\over 2 m_{J/\psi}}\,,
\eea
where we use $\vec q=  (1,1,1)\times \theta_0/L$. For that purpose, and for each of our lattices, we first computed the masses of $m_{J/\psi}$ and of $m_{\eta_c}$, and then by using eq.~(\ref{eq:angle1}) we determined $\theta_0$ that is then used in the computation of one of the charm quark propagators. In practice this last step is made by ``twisting" the gauge links according to
\bea
U_\mu(x) \to U_\mu^\theta=e^{i\theta_\mu/L}U_\mu(x)\,, ~~ {\rm where}\; \theta_\mu=(0,\vec \theta)\,,
\eea
on which the quark propagator is computed according to,
\bea\label{eq:twisted-prop}
S_c^{\vec \theta}(x,0;U)=e^{i\vec \theta\cdot \vec x/L}S_c(x,0;U^\theta)\,.
\eea
In our notation the quark propagator $S_c(x,0) \equiv S_c(\vec x,t; \vec 0, 0) = \langle \bar c(x)c(0)\rangle_U$, and we only in eq.~(\ref{eq:twisted-prop}) we write explicitly the gauge field configuration in the argument, to distinguish $U$ from $U^\theta$. In what follows $U$ will be implicit.

Similarly, in the computation of the physical $h_c\to \eta_c\gamma$ decay, we compute the transition matrix element that is parameterized in terms of two form factors as,~\footnote{From now on we will drop the label $1P$, and write $h_c$ only.} 
\bea\label{def-vectorFH}
\langle \eta_c(k) \vert  J^{\rm em}_\mu \vert h_c(p,\epsilon_\lambda) \rangle &=& 
i e {\cal Q}_c\ \left\{ m_{h_c} F_1(q^2) \left(  \epsilon^{\lambda \ast}_\mu - { \epsilon_\lambda^\ast  \cdot q\over q^2} q_\mu \right)\right.  \nn\\
&&\left. \qquad + F_2(q^2) ( \epsilon_\lambda^\ast \cdot q) \left[ 
 { m_{h_c}^2 - m_{\eta_c}^2\over q^2} q_\mu  -  (p+k)_\mu \right]\right\}\,. 
\eea
The decay rate for the on-shell photon is~\cite{lattice-radiative-1}
\bea\label{widthH}
\Gamma(h_c\to \eta_c\gamma)= {16\over 27}   {    \alpha_{\rm em}    \vert  \  \vec   q \  \vert} \cdot  \left| F_1(0)\right|^2= {8\over 27}\ \alpha_{\rm em}\  { m_{h_c}^2-m_{\eta_c}^2 \over  m_{h_c}} \ \left| F_1(0)\right|^2\,.
\eea
To reach the physical form factor at $q^2=0$, with $h_c$ at rest, the twisted boundary condition applied on one of the charm quark propagators is made with 
\bea\label{eq:angle2}
\tilde \theta_0 = {L\over \sqrt{3}} {m_{h_c}^2-m_{\eta_c}^2\over 2 m_{h_c}}\,.
\eea

\section{Two-point correlation functions}

\begin{figure}[t!]
\begin{center}
\begin{tabular}{@{\hspace{-0.6cm}}c}
\epsfxsize9.6cm\epsffile{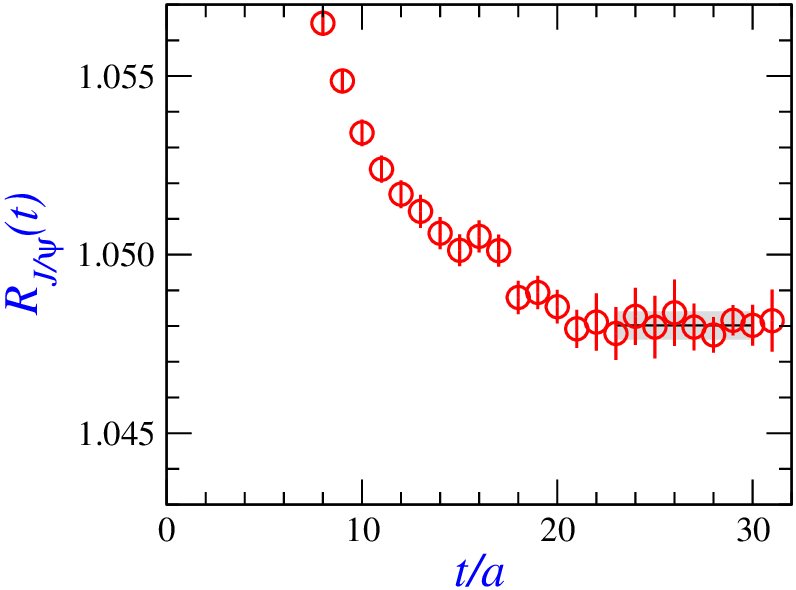}   \\
\end{tabular}
\vspace*{-.1cm}
\caption{\label{fig:10}{\footnotesize 
Plateau of $R_{J/\psi}(t)$, defined in eq.~(\ref{eq:RRR}), obtained from our computations at $(\beta,\mu_{\rm sea})=(4.2,0.0065)$. The shaded area is the result of the fit to a constant  $R_{J/\psi}$. Results for all other values of $(\beta,\mu_{\rm sea})$ explored in this study are listed in tab.~\ref{tab:2}. 
 } }
\end{center}
\end{figure}

\begin{table}[h!!]
\centering 
{\scalebox{.93}{\begin{tabular}{|c|cccccc|}  \hline \hline
{\phantom{\huge{l}}}\raisebox{-.2cm}{\phantom{\Huge{j}}}
$ \beta$& 3.8 &  3.9  &  3.9 & 4.05 & 4.2  & 4.2    \\ 
{\phantom{\huge{l}}}\raisebox{-.2cm}{\phantom{\Huge{j}}}
$ L^3 \times T $&  $24^3 \times 48$ & $24^3 \times 48$  & $32^3 \times 64$ & $32^3 \times 64$& $32^3 \times 64$  & $48^3 \times 96$  \\ 
{\phantom{\huge{l}}}\raisebox{-.2cm}{\phantom{\Huge{j}}}
$ \#\ {\rm meas.}$& 240 &  240  & 150  & 150 & 150 & 100  \\ \hline 
{\phantom{\huge{l}}}\raisebox{-.2cm}{\phantom{\Huge{j}}}
$\mu_{\rm sea 1}$& 0.0080 & 0.0040 & 0.0030 & 0.0030 & 0.0065 &  0.0020   \\ 
{\phantom{\huge{l}}}\raisebox{-.2cm}{\phantom{\Huge{j}}}
$\mu_{\rm sea 2}$& 0.0110 & 0.0064 & 0.0040 & 0.0060 &   &     \\ 
{\phantom{\huge{l}}}\raisebox{-.2cm}{\phantom{\Huge{j}}}
$\mu_{\rm sea 3}$&  & 0.0085 &  & 0.0080 &   &     \\ 
{\phantom{\huge{l}}}\raisebox{-.2cm}{\phantom{\Huge{j}}}
$\mu_{\rm sea 4}$&  & 0.0100 &  &   &   &     \\   \hline 
{\phantom{\huge{l}}}\raisebox{-.2cm}{\phantom{\Huge{j}}}
$a \ {\rm [fm]}$&   0.098(3) & 0.085(3) & 0.085(3) & 0.067(2) & 0.054(1) & 0.054(1)      \\ 
{\phantom{\huge{l}}}\raisebox{-.2cm}{\phantom{\Huge{j}}}
$Z_V (g_0^2)$~\cite{ZZZ}& 0.5816(2) & 0.6103(3) & 0.6103(3)  & 0.6451(3) & 0.686(1) & 0.686(1) \\ 
{\phantom{\huge{l}}}\raisebox{-.2cm}{\phantom{\Huge{j}}}
$Z_A (g_0^2)$~\cite{ZZZ}& 0.746(11) & 0.746(6) & 0.746(6)  & 0.772(6) & 0.780(6) & 0.780(6) \\ 
{\phantom{\huge{l}}}\raisebox{-.2cm}{\phantom{\Huge{j}}}
$\mu_{c}$~\cite{Blossier:2010cr}& 0.2331(82)  &0.2150(75)  &0.2150(75)   & 0.1849(65) & 0.1566(55) & 0.1566(55)  \\ 
 \hline \hline
\end{tabular}}}
{\caption{\footnotesize  \label{tab:1} Summary of the details about the lattice ensembles used in this work (for more information see ref.~\cite{boucaud}).  Data obtained at different $\beta$'s are rescaled by using the Sommer parameter $r_0/a$, and the overall lattice spacing is fixed by matching $f_\pi$ obtained on the lattice with its physical value, leading to  $r_0= 0.440(12)$~fm (c.f. ref.~\cite{Blossier:2010cr}). All quark masses are given in lattice units.}}
\end{table}
Similarly to our recent publication~\cite{Becirevic:2012ti}, we use the gauge field configurations produced by ETM Collaboration~\cite{boucaud} employing the maximally twisted mass QCD~\cite{fr}, the details of which are summarized in tab.~\ref{tab:1}. The masses of charmonia are extracted from the following correlation functions:
\bea\label{eq:0}
 C^{\eta_c}(t)  &=&    \langle  \sum_{\vec x }{\rm Tr}\left[ S_c(0,0;\vec x,t)\gamma_5S_c^\prime (\vec x,t;\vec 0,0) \gamma_5\right] \rangle\,,\nn\\
 C^{J/\psi}_{ii}(t) &=&   {1\over 3}\sum_{i=1}^3  \langle  \sum_{\vec x }{\rm Tr}\left[ S_c(0,0;\vec x,t)\gamma_i S_c^\prime(\vec x,t;\vec 0,0) \gamma_i\right] \rangle\,,\nn\\
 C_{ij}^{h_c}(t) &=&   {1\over 3}\sum_{i,j=1}^3  \langle  \sum_{\vec x }{\rm Tr}\left[ S_c(0,0;\vec x,t)\gamma_i \gamma_jS_c^\prime(\vec x,t;\vec 0,0)\gamma_i \gamma_j\right] \rangle_{i\neq j},
\eea
in which the Dirac structures are chosen to provide the coupling to the charmonium states with quantum numbers $J^{PC}=0^{-+}$, $1^{--}$, and $1^{+-}$, for $\eta_c$, $J/\psi$ and $h_c$, respectively.  
$S_c(0,0;\vec x,t)$ and $S_c^\prime(0,0;\vec x,t)$ refer to the propagators of the charm quark in the doublet $\psi(x) = [c(x)\ c^\prime(x)]^T$ entering the maximally twisted mass QCD action on the lattice~\cite{fr}~\footnote{Note that we write the action in the ``physical basis" and not in the twisted one. }
\bea
S=a^4\sum_{x}\bar \psi(x)\left\{ \frac{1}{2}\sum_\mu\gamma_\mu\left(\nabla_\mu+\nabla_\mu^\ast\right)-i\gamma_5\tau^3 r \left[m_{\rm cr}-\frac{a}{2}\sum_\mu\nabla_\mu^\ast\nabla_\mu\right]+\mu_c\right\}\psi(x)\,,
\eea
and therefore the propagator $S_c(0,0;\vec x,t)$ is obtained by inverting the above lattice Dirac operator with the Wilson parameter $r$, while $S_c^\prime(0,0;\vec x,t)$  is obtained by using $-r$. In practice $r=1$ and $m_{\rm cr}$ is the same as the one used in the production of the gauge field configurations~\cite{boucaud}. Finally, $\nabla_\mu$ and $\nabla_\mu^\ast$ are the usual forward and backward derivatives on the lattice. 
In this study we also implement the Gaussian smearing on one of the sources~\cite{gattringer}. In other words, one replaces $c(x)$ by
\begin{equation}
c_{n_g}=\left(\frac{1+\kappa H}{1+6\kappa}\right)^{n_g}c\,,
\end{equation}
where the smearing operator $H$ is defined via~\cite{Gusken:1989ad}
\begin{equation}
H_{i,j}=\sum_{\mu=1}^3\left(U^{n_a}_{i;\mu}\delta_{i+\mu,j}+U^{n_a\dagger}_{i-\mu;\mu}\delta_{i-\mu,j}\right)\,,
\end{equation}
where $U^{n_a}_{i,\mu}$ is the $n_a$ times APE smeared link~\cite{albanese}, defined in terms of $(n_a-1)$ times smeared link $U^{(n_a-1)}_{i,\mu}$ and its surrounding staples $V^{(n_a-1)}_{i,\mu}$, 
\begin{equation}
U^{n_a}_{i,\mu}={\rm Proj_{SU(3)}}\left[(1-\alpha)U^{(n_a-1)}_{i,\mu}+\frac{\alpha}{6} V^{(n_a-1)}_{i,\mu}\right ]\,.
\end{equation}
We chose the parameters
\bea
\kappa=4,\,n_g=30,\,\alpha=0.5,\,n_a=20\,,
\eea
which are kept fixed for all of our lattices.  
The value of the bare charm quark mass, $\mu_c$, at each of our lattices is given in tab.~\ref{tab:1}.  It has been fixed according to the result of ref.~\cite{Blossier:2010cr} where it was shown that the charm quark computed from the comparison of the lattice results with the physical $m_{\eta_c}$ fully agrees with the value obtained by using the physical $m_{D_s}$ or $m_{D}$. Therefore,  we can say that $m_{\eta_c}$, obtained by numerically solving
\begin{figure}[t!!]
\begin{center}
\begin{tabular}{@{\hspace{-0.25cm}}c}
\epsfxsize8.2cm\epsffile{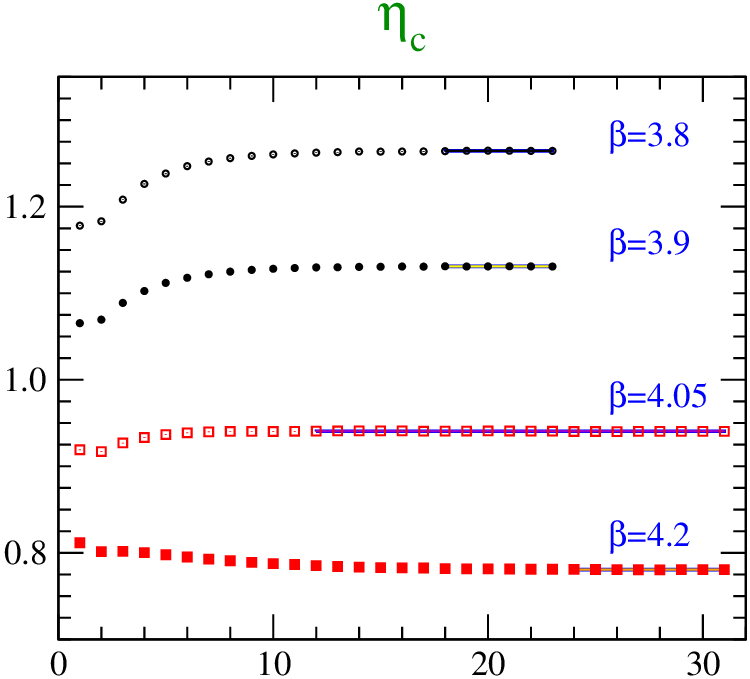}~\epsfxsize8.2cm\epsffile{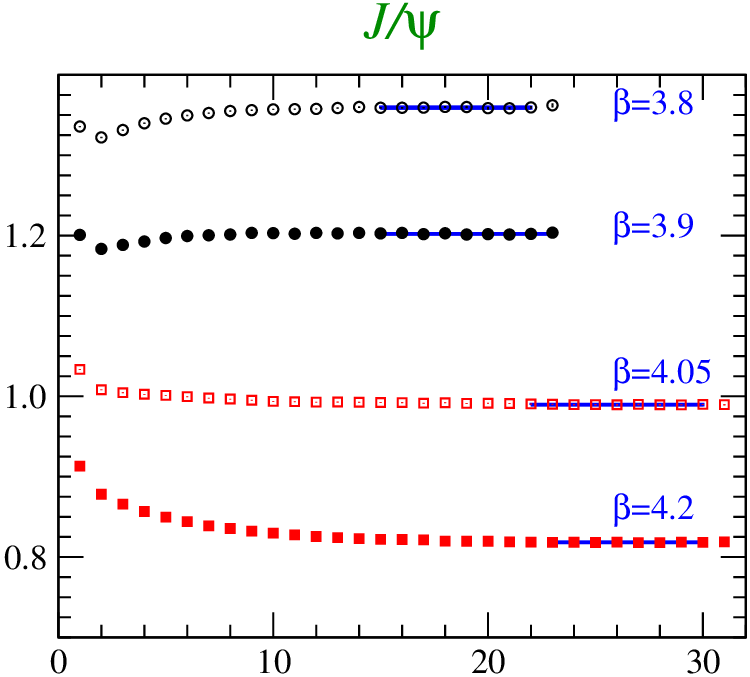}    \\
\epsfxsize8.2cm\epsffile{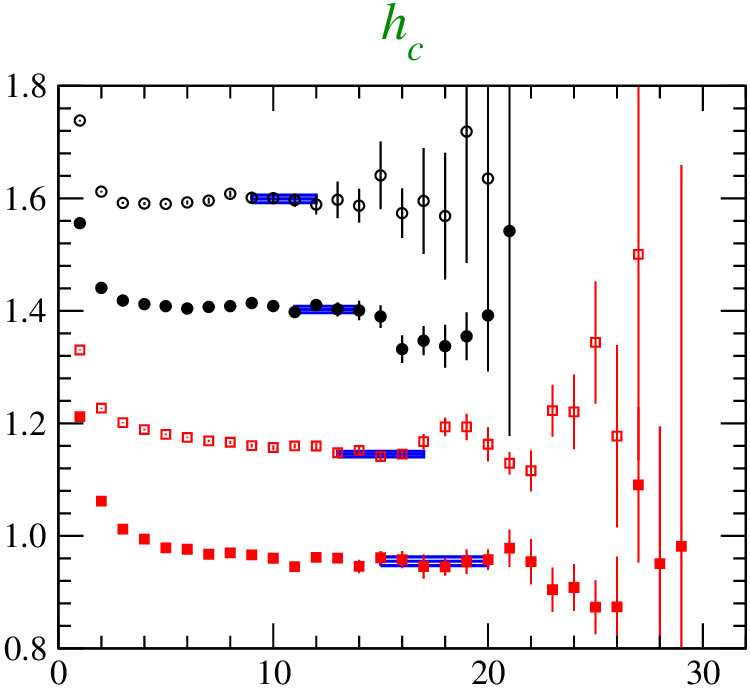}   \\
\end{tabular}
\vspace*{-.1cm}
\caption{\label{fig:1}{\footnotesize 
Effective masses of the charmonium states, $m_{\eta_c,J/\psi,h_c}^{\rm eff}(t)$, extracted from the two-point correlation functions according to eqs.~(\ref{eta-eff},\ref{psi-eff}) at four lattice spacings.  Illustration is  provided for one value of the sea quark mass. Note also that the smearing parameters used in this work are kept fixed to the same values for all our lattices. }}
\end{center}
\end{figure}
\bea\label{eta-eff}
{\cosh\left[ m_{\eta_c}^{\rm eff}(t) \left( {\displaystyle{T\over 2}} - t\right)\right] \over \cosh\left[ m_{\eta_c}^{\rm eff}(t) \left( {\displaystyle{T\over 2}} - t -1\right)\right] }  = {C^{{\eta_c}}(t)\over C^{\eta_c}(t+1)}\,,
\eea  
and then by fitting $m_{\eta_c}^{\rm eff}(t)$ at large time separations to a constant, is merely  a verification that, after a smooth continuum extrapolation, we indeed reproduce $m_{\eta_c}^{\rm exp.}=2.980(1)$~GeV. To extract the values of 
$m_{J/\psi}$ and $m_{h_c}$ we proceed along the same line and compute  $m_{J/\psi,h_c}^{\rm eff}(t)$ from
\bea\label{psi-eff}
{\cosh\left[ m_{J/\psi}^{\rm eff}(t) \left( {\displaystyle{T\over 2}} - t\right)\right] \over \cosh\left[ m_{J/\psi}^{\rm eff}(t) \left( {\displaystyle{T\over 2}} - t -1\right)\right] }  = {C_{ii}^{{J/\psi}}(t)\over C_{ii}^{J/\psi}(t+1)}\,,\nn\\
&&\hfill \nn\\
&&\hfill \nn\\
{\cosh\left[ m_{h_c}^{\rm eff}(t) \left( {\displaystyle{T\over 2}} - t\right)\right] \over \cosh\left[ m_{h_c}^{\rm eff}(t) \left( {\displaystyle{T\over 2}} - t -1\right)\right] }  = {C_{ij}^{{h_c}}(t)\over C_{ij}^{h_c}(t+1)}\,.
\eea
In fig.~\ref{fig:1} we show all three effective mass plots as obtained by using all four lattice spacings explored in this work and for one value of the sea quark mass which we choose to be the least light ones. We see that the effective masses for the pseudoscalar ad vector charmonia are excellent while the signal for the orbitally excited state, $h_c$, is much more noisy.  The effective masses are then combined to
\bea\label{eq:RRR}
R_{J/\psi}(t)= {m_{J/\psi}^{\rm eff}(t)\over m_{\eta_c}^{\rm eff}(t)}\,,\quad 
R_{h_c}(t)= {m_{h_c}^{\rm eff}(t)\over m_{\eta_c}^{\rm eff}(t)}\,.
\eea
The advantage of these ratios is that they have smaller statistical errors  than any of the effective meson masses separately. In fig.~\ref{fig:10} we show one such a ratio. On the plateaus, that we carefully examined for each of our lattices, we then fit $R_{J/\psi,h_c}(t)$ to a constant $R_{J/\psi,h_c}$.  In tab.~\ref{tab:2} we collect our results for $R_{J/\psi,h_c}$ as obtained from all of the lattice ensembles at our disposal. As indicated in the plot in fig.~\ref{fig:1} the fitting intervals involving the state  $h_c$ are shorter. The shift of the plateau region to the right is made to account for the different lattice spacings, so that the fit is made at approximately the same physical separation between the interpolating field operators. 

After a smooth linear extrapolation to the continuum limit, 
\bea\label{eq:contR}
R_{J/\psi,h_c} = R_{J/\psi,h_c}^{\rm cont.} \left[ 1 +  b_{J/\psi,h_c} m_{q} + c_{J/\psi,h_c} {a^2\over (0.086\ {\rm fm})^2}\right]\,.
\eea
we obtain
\bea\label{eq:resR}
R_{J/\psi }^{\rm cont.}  = 1.0377(6)&&\quad[\ 1.0391(4)\ ]^{\rm exp.}\,,\cr
R_{ h_c}^{\rm cont.} = 1.187(11)&&\quad [\ 1.1829(5)\ ]^{\rm exp.}\,.
\eea
For the reader's convenience we also quoted the values obtained from experiments~\cite{PDG}.  In eq.~(\ref{eq:contR}) the parameter $b_{J/\psi,h_c}\approx 0$ measures the dependence on the sea quark mass, $m_q\equiv m_q^\msbar(2\ \gev)$, while the parameter $c_{J/\psi,h_c}\approx 3$~\% measures the leading discretization effects. Division by $a_{\beta=3.9}=0.086$~fm is made for convenience. The linear fit~(\ref{eq:contR}) describes our data very well except for the results obtained at $\beta=3.8$. The results obtained at $\beta=3.8$ can be either excluded from the continuum extrapolation, which is how we got the above results, or a term proportional to $a^4$ can be added in (\ref{eq:contR}) which leads to a result that is fully consistent with the one quoted above.  Finally, we should stress that the disconnected, OZI-suppressed, contributions to the correlation functions discussed in this work have been neglected. The fact that our lattice results agree with the experimental values (\ref{eq:resR}) can be viewed as a proof that the OZI suppressed contributions to the two-point functions are indeed very small.

\subsection{Dispersion Relation and the Sea Quark Mass Dependence}

We already mentioned that for the determination of the desired radiative decay form factors one of the charm quark propagators is to be computed with twisted boundary conditions~(\ref{eq:twisted-prop}), with the twisting angle tuned to ensure $q^2=0$, c.f. eqs.~(\ref{eq:angle1},\ref{eq:angle2}). In this section we check on the energy-momentum relation in the case of the pseudoscalar charmonium $\eta_c$ by exploring five different values of the twisting angle, covering the range of the meson's three-momenta, $0\leq |\vec p| \leq \sqrt{2}\times 2\pi/L$. We compute
\bea
 C^{\eta_c}(E_p;t)  &=&   \langle  \sum_{\vec x }{\rm Tr}\left[ S_c^{\vec \theta}(0,0;\vec x,t)\gamma_5S_c^\prime (\vec x,t;\vec 0,0) \gamma_5\right] \rangle\,,
\eea
with $\vec p=\vec \theta/L$. For $|\vec \theta |=0$ we obviously get $m_{\eta_c}$. For  $|\vec \theta |\neq 0$, we proceed like in eq.~(\ref{eta-eff}) and fit $E^{\rm eff}_p(t)$ to a constant $E_p$.
\begin{figure}[h!!]
\begin{center}
\begin{tabular}{@{\hspace{-0.6cm}}c}
\epsfxsize9.2cm\epsffile{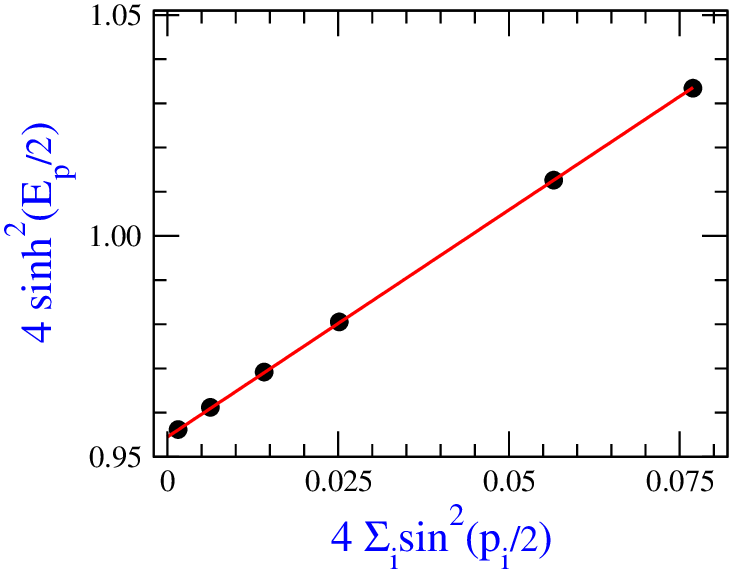}   \\
\end{tabular}
\vspace*{-.1cm}
\caption{\label{fig:2}{\footnotesize 
Test of the free boson lattice dispersion relation~(\ref{eq:freeboson}) on our results obtained at $\beta =4.05$ in the case of the pseudoscalar charmonium $\eta_c$. 
 } }
\end{center}
\end{figure}
As expected, due to the discretization effects the continuum relativistic formula, $E_p^2 = m_{\eta_c}^2 + \vec p^{\ 2}$, is not accurately verified on the lattice. Instead, our results satisfy 
\bea\label{eq:freeboson}
4\sinh^2{E_p\over 2} = 4 \sin^2{|\vec p|\over 2} +  C_\beta\ 4 \sinh^2{m_{\eta_c}\over 2}\,,
\eea
rather well. For $C_\beta=1$ the above formula is the dispersion relation of a free boson on the lattice. After fitting our results to the above expression we find, for each of our lattice spacings, 
\bea
C_\beta = 0.97(1)_{3.8},1.00(1)_{3.9},1.03(1)_{4.05},1.04(2)_{4.2}\,.
\eea
The results at $\beta =4.05$ are shown in fig.~\ref{fig:2}.  At this stage it is not clear whether or not with higher statistics all of the above $C_\beta$ values would get closer to $1$. The finite volume effects on the sea quark mass are unlikely to modify the expression~(\ref{eq:freeboson}). That point we could check from our simulations at $\beta=3.9$ and $\mu_{\rm sea}=0.0040$ where the results obtained at two lattices $24^3\times 48$ and $32^3\times 64$ are perfectly consistent.

\begin{figure}[h!!]
\begin{center}
\begin{tabular}{@{\hspace{-0.6cm}}c}
\epsfxsize8.2cm\epsffile{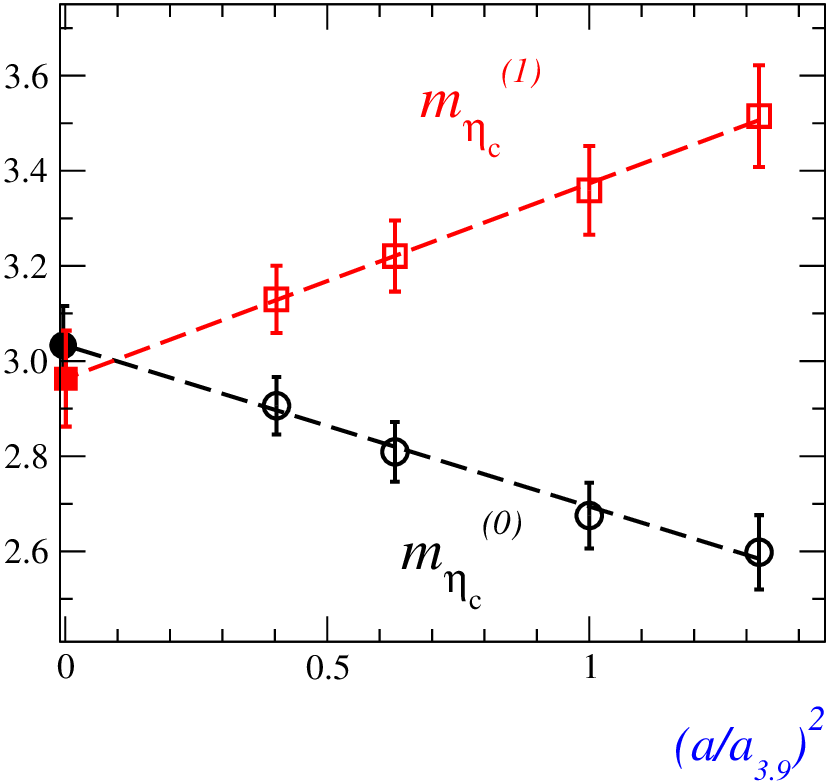}    \\
\end{tabular}
\vspace*{-.1cm}
\caption{\label{fig:3}{\footnotesize 
$m_{\eta_c}^{(0)}$ and  $m_{\eta_c}^{(1)}$ are the so called rest and kinetic mass obtained from the fit of our data to eq.~(\ref{eq:fnal}). The above plot shows extrapolation of their values (in physical units, GeV) to the continuum limit. } }
\end{center}
\end{figure}

Since we are working with heavy quarks it is tempting to use the non-relativistic energy-momentum relation as well, but accounting for the lattice artifacts that according to ref.~\cite{ElKhadra:1996mp} can be included by the distinction between the ``rest" ($m_{\eta_c}^{(0)}$) and the ``kinetic'' mass ($m_{\eta_c}^{(1)}$) in,
\bea\label{eq:fnal}
E(\vec p) =  m_{\eta_c}^{(0)} + { \vec p^{\ 2}\over 2 m_{\eta_c}^{(1)} } + \dots
\eea
We fit our data to this expression and find that at each lattice spacing the kinetic mass is indeed larger than the rest one but they ultimately converge to the same value in the continuum limit, as they should on the basis of the restored Lorentz invariance, as shown in fig.~\ref{fig:3}. 
Interestingly, however, we see that the discretization error in both the rest and kinetic masses are of the same size, but of the opposite sign.

The above discussion on the dispersion relations is made on the data obtained at fixed value of the sea quark mass. We checked that indeed
\bea
{\partial m_{\eta_c}\over \partial m_{\rm sea}}=0\,.
\eea
The same is, however, not true with $m_{J/\psi}$, although this is hard to see from the mass ratio in eq.~(\ref{eq:contR}) alone. 
Instead we examined the hyperfine splitting, 
\bea
\Delta = m_{J/\psi}-m_{\eta_c} = m_{\eta_c}( R_{J/\psi}-1) \,, 
\eea
and from the fit of our data to, 
\bea\label{eq:fitdelta}
\Delta = \Delta^{\rm cont.} \left[ 1 +  b_{\Delta} m_{q} + c_{\Delta} {a^2\over (0.086\ {\rm fm})^2}\right]\,,
\eea
\begin{figure}
\vspace*{-0.8cm}
\begin{center}
\begin{tabular}{@{\hspace{-0.25cm}}c}
\epsfxsize10.7cm\epsffile{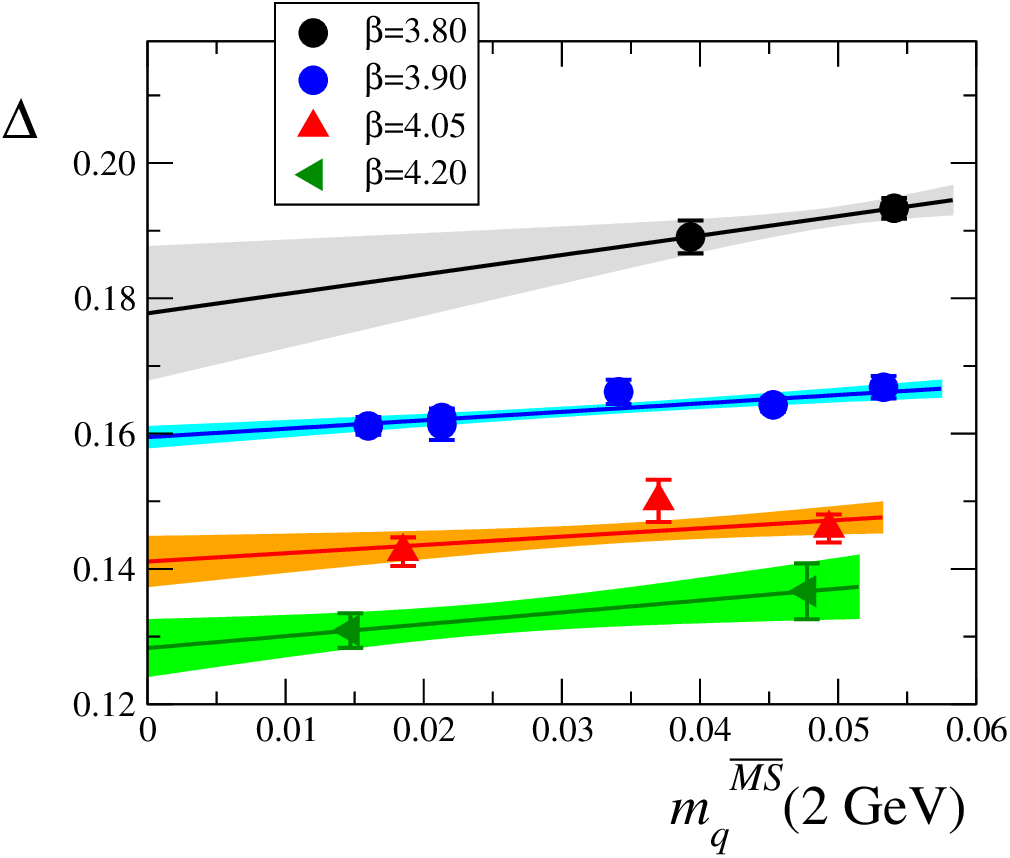}    \\
\end{tabular}
\vspace*{-.1cm}
\caption{\label{fig:4}{\footnotesize 
Hyperfine splitting in charmonium as a function of the light sea quark mass [$m_{\rm sea}\equiv m_q^\msbar(2\ \gev)$]. Four lines correspond to the fit  $\Delta = \Delta^{(0)} \left[ 1 +  \bar b_{\Delta} m_{q}\right]$ made at each of our lattice spacings separately. All quantities are given in physical units [GeV].}}
\end{center}
\end{figure}
we find  
\bea\label{latt-D}
\Delta^{\rm cont.} = (112\pm 4)~\mev&&\quad[\ 116.6\pm 1.2\ ]^{\rm exp.}\,,
\eea
in good agreement with the experimental result written in brackets~\cite{PDG}, and in excellent agreement with the result of lattice QCD computation presented in ref.~\cite{follana}, $\Delta = (111\pm 5)~\mev$. Once more, this implicitly suggests that the contribution of the OZI breaking diagrams in the charmonia,  neglected in our computations, are very small (c.f. discussion in ref.~\cite{disconn}). 
Note also that from the fit of our data to eq.~(\ref{eq:fitdelta}) we find, 
\bea
b_\Delta =1.0(3)\ \gev^{-1}\,,\quad c_\Delta= 0.47(6)\,,
\eea
in qualitative agreement with ref.~\cite{0912} where a tiny decrease of $\Delta$ is found while lowering the sea quark mass. Note, however, that this observation ($b_\Delta \gtrapprox 0$) disagrees with earlier findings of ref.~\cite{Manke:2000dg}.  To better appreciate that disagreement we fit our data at each lattice spacing separately as a linear function of the light sea quark mass and plot the resulting curves in fig.~\ref{fig:4}.  We see that at each of our $\beta$'s the hyperfine splitting mildly decrease as the sea quark mass approaches the chiral limit. 
The above value for $\Delta^{\rm cont.}$ is obtained without including the results at $\beta=3.8$ in the continuum extrapolation. If they are included our final result does not change but the error becomes smaller by $1~\mev$. To be consistent with what we quote as a result of the continuum extrapolation in all other quantities computed in this work, we will quote $\Delta = (112\pm 4)~\mev$.

\section{\label{sec-3}Radiative Transition Form Factors}

\begin{figure}
\vspace*{-0.8cm}
\begin{center}
\begin{tabular}{@{\hspace{-0.25cm}}cc}
\epsfxsize7.9cm\epsffile{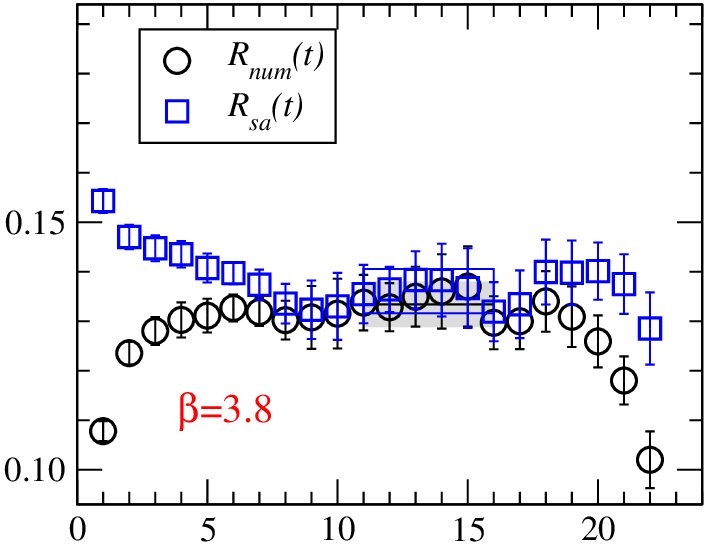}&\epsfxsize7.9cm\epsffile{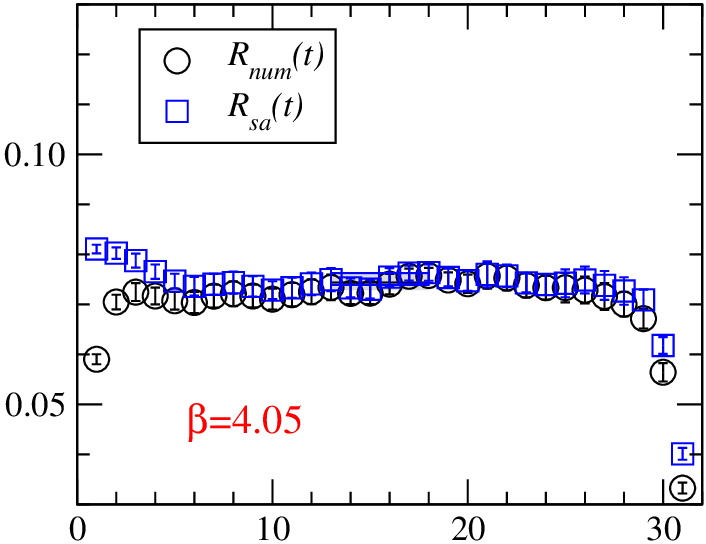}    \\
\epsfxsize7.9cm\epsffile{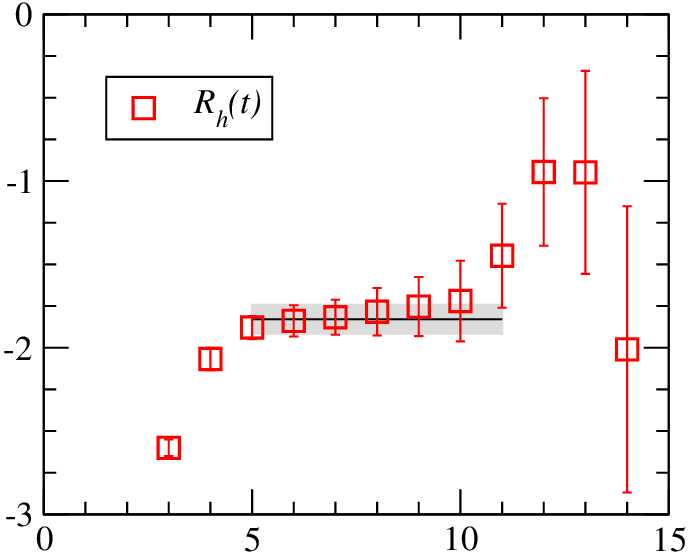}&\epsfxsize7.9cm\epsffile{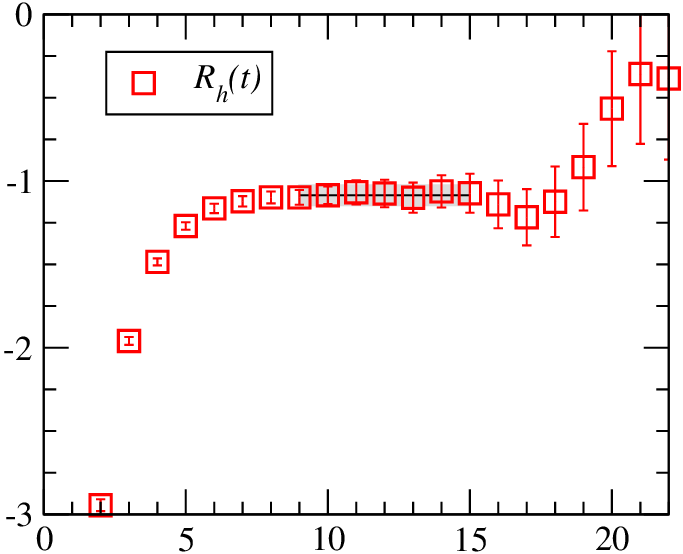}    \\
\end{tabular}
\vspace*{-.1cm}
\caption{\label{fig:5}{\footnotesize 
Plateaus exhibited by the semi-analytic and numerical ratios, $R_{\rm sa, num}(t)$ defined in eqs.~(\ref{eq:rnum},\ref{eq:rsa}), are shown in the upper plots, while on the lower ones we show the semi-analytic ratios $R_{\rm h}(t)$~(\ref{eq:Rh}). Since our results do not depend on the mass 
of the light sea quark, we plot the plateaus for the largest value of the sea quark mass at two values of $\beta$. 
 } }
\end{center}
\end{figure}

To extract the desired hadronic matrix element~(\ref{def-vectorFF}) we computed the three point correlation functions
\bea\label{eq:c31}
C_{ij}(\vec q;t)&=&  \sum_{\vec x,\vec y} \langle  V_i^\dag (0) J^{\rm em}_j(x) P(y)  \rangle \  e^{i\vec q\cdot (\vec x -\vec y )}  \nn\\
  &=&   \langle \sum_{\vec x,\vec y}{\rm Tr}\left[  S_c^\prime (y;0) \gamma_i S_c (0,x)\gamma_j   S_c^{ \vec \theta}(x,y) \gamma_5 \right]\rangle \,,
\eea
where $P=\bar c\gamma_5 c^\prime$,  $V_i =\bar c\gamma_i c^\prime$ are the interpolating operators fixed at $t=0$ and $t=t_y=T/2$ ($T$ being the time extension of our lattices). Using the fact that our three-momentum is isotropic, $\vec q = -(1,1,1)\times \theta_0/L$,  we can average
\bea\label{eq:vvv}
C_V(\vec q;t)&=& {1\over 6} \left[
C_{12}(\vec q;t) + C_{23}(\vec q;t) + C_{31}(\vec q;t) - C_{21}(\vec q;t)- C_{32}(\vec q;t) - C_{13}(\vec q;t)\right]\nn\\
&\to& {{\cal Z}_P^S  \over 2 E_{\eta_c}}e^{-E_{\eta_c} (t_{\rm fix} - t)} \times {\theta_0\over L} {2 m_{J/\psi}\over m_{J/\psi} + m_{\eta_c}} V(0)
\times  {  {\cal Z}_V^S\over 2 m_{J/\psi} }e^{- m_{J/\psi} t}
\,,
\eea
where the last line is valid for the sufficiently separated operators in the correlation function~(\ref{eq:c31}), which is ensured by the Gaussian smearing procedure. Coupling to the smeared pseudoscalar (vector) interpolating field operator $P$ ($V$), is denoted by  ${\cal Z}_P^S$ (${\cal Z}_V^S$).~\footnote{The values of ${\cal Z}_P^S$ and ${\cal Z}_V^S$ are easily computed from the correlators of smeared source operators as
\bea
C^{\eta_c}(t) \to \left(|{\cal Z}_P^S|^2/2 m_{\eta_c}\right) \exp[-m_{\eta_c} t] \,, \qquad C^{J/\psi}_{ii}(t) \to  \left(|{\cal Z}_V^S|^2/2 m_{J/\psi}\right) \exp[- m_{J/\psi}t] \,,
\nn\eea
to which we include the signal propagating from the opposite end of our lattice.} Electromagnetic current, $J^{\rm em}_j = Z_V(g_0^2) \bar c\gamma_j c$, is  local and renormalized by using $Z_V(g_0^2)$ listed in tab.~\ref{tab:1}. Notice also that the pseudoscalar source operator has been fixed at $t_{y}=T/2$, which simplifies the averaging of the signals propagating in both halves of the lattice, i.e. $t\in (0,\pm T/2)$.  In computing the propagators $S_c(x;y)$ we used the stochastic source techniques as explained in ref.~\cite{boucaud}. Since the signals for the pseudoscalar and vector charmonia are very good, we can proceed to eliminate the sources in two ways. We can either divide the correlator~(\ref{eq:c31}) by the corresponding two point functions, 
\bea\label{eq:rnum}
R_{\rm num}(t) = {C_V(\vec q;t) \over C_{ii}^{J/\psi}(t)  C^{\eta_c}(E_q;T/2-t) } \times {\cal Z}_P^S {\cal Z}_V^S\,,
\eea
which we refer to as the {\sl numerical ratio}, or by the analytic expression of the coupling of the source operators to the lowest states, 
\bea\label{eq:rsa}
R_{\rm sa}(t) = {C_V(\vec q;t) \over {\cal Z}_P^S {\cal Z}_V^S} \times 4 m_{J/\psi}  E_{\eta_c}\  e^{m_{J/\psi} t + E_{\eta_c} (T/2 - t)}\,,
\eea
which we call the {\sl semi-analytic} ratio. Obviously, the values for ${\cal Z}_{P,V}^S$, $m_{J/\psi}$ and $E_{\eta_c}$ are obtained from the study of the two-point correlation functions. The resulting plateaus for all the lattice spacings used in this work are shown in fig.~\ref{fig:5}.
We see that thanks to efficiency of the smearing procedure the plateaus of the two ratios indeed coincide over a broad range of time-slices. In the plots shown in fig.~\ref{fig:5}  we also indicate the intervals (shaded areas) in which we fitted the ratios $R_{\rm sa, num}(t)$ to a constant. This is then used to obtain the form factor $V(0)$, as indicated in eq.~(\ref{eq:vvv}). 

\begin{figure}[h]
\begin{center}
\begin{tabular}{@{\hspace{-0.25cm}}c}
\epsfxsize8.2cm\epsffile{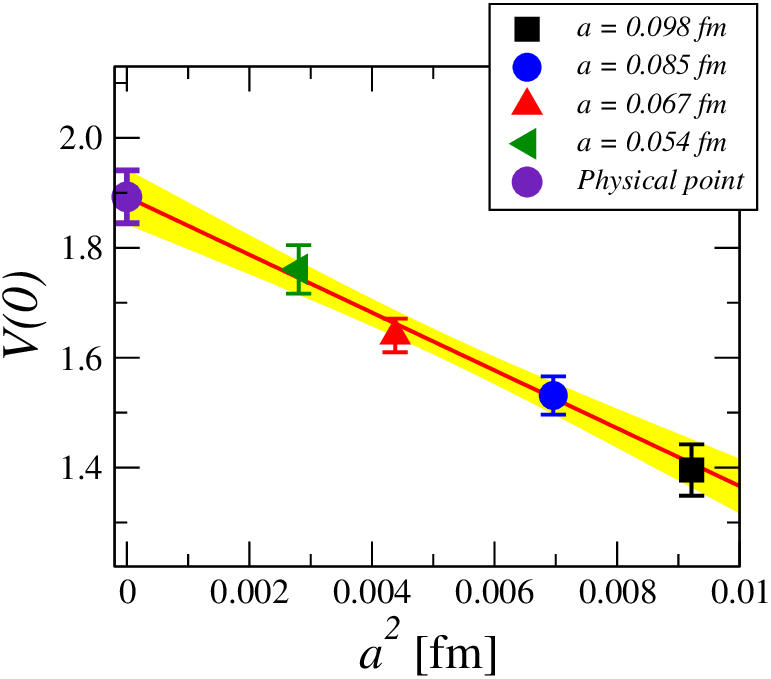}~\epsfxsize8.2cm\epsffile{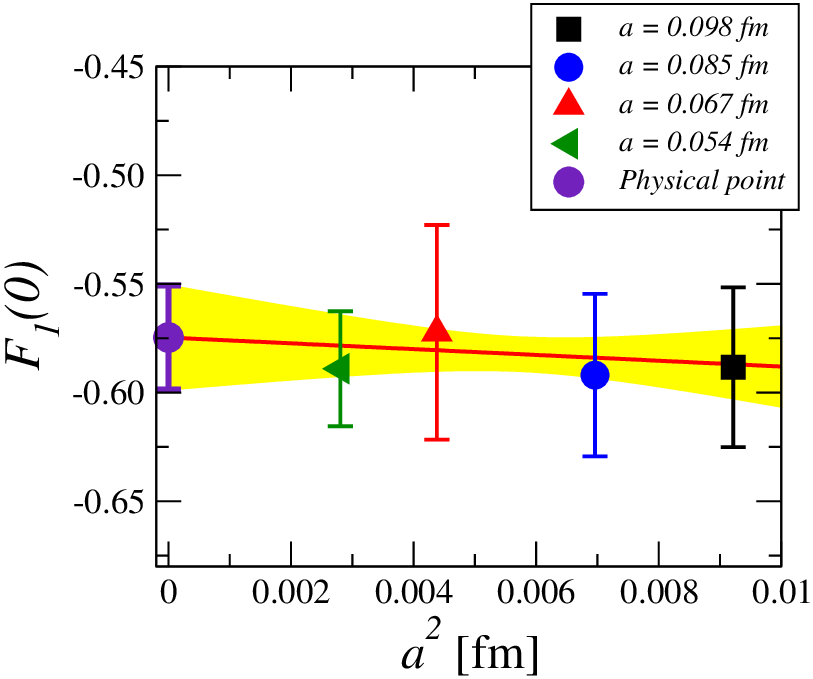}   \\
\end{tabular}
\vspace*{-.1cm}
\caption{\label{fig:6}{\footnotesize 
Continuum extrapolation of the form factors $V(0)$ and $F_1(0)$ computed on our lattices at $4$ lattice spacings. Note that for every value of $a^2$ we have several data points obtained for different value of the light sea quark mass. The continuum extrapolation is made according to eq.~(\ref{extrap-V}), and similarly for the form factor $F_1(0)$.
 } }
\end{center}
\end{figure}

In tab.~\ref{tab:2} we present our results for the form factor $V(0)$, as obtained from all of the lattice data-sets and by using the ratio $R_{\rm sa}(t)$.  
\begin{table}[h!!]
\centering 
{\scalebox{.93}{\begin{tabular}{|cc|cccccc|}  \hline 
{\phantom{\huge{l}}}\raisebox{-.3cm}{\phantom{\Huge{j}}}
$(\beta,\mu_{\rm sea})$ & $\; L\quad $ &  $a m_{\eta_c}$  & $R_{J/\psi}$ & $R_{h_c}$ & $a f_{J/\psi}$ & $V(0)$  & $F_1(0)$    \\    \hline 	 \hline 	
{\phantom{\huge{l}}}\raisebox{-.2cm}{\phantom{\Huge{j}}}
(3.80, 0.0080) & 24 & 1.2641(2) & 1.0749(6) & 1.254(5) &  0.226(1) &  1.36(6) &  -0.58(3) \\			
{\phantom{\huge{l}}}\raisebox{-.2cm}{\phantom{\Huge{j}}}                            
(3.80, 0.0110) & 24 & 1.2645(3) & 1.0749(4) & 1.265(5)  &  0.224(2)&  1.42(5) &  -0.59(4) \\ \hline 
{\phantom{\huge{l}}}\raisebox{-.2cm}{\phantom{\Huge{j}}}                                      
(3.90, 0.0040) & 24 & 1.1308(4) & 1.0621(5) & 1.235(6)  &  0.184(2)&  1.48(4) &  -0.60(4) \\	            
{\phantom{\huge{l}}}\raisebox{-.2cm}{\phantom{\Huge{j}}}                                      
(3.90, 0.0064) & 24 & 1.1311(2) & 1.0628(4) & 1.235(6)  &  0.186(2)&  1.48(3) &  -0.56(2) \\	            
{\phantom{\huge{l}}}\raisebox{-.2cm}{\phantom{\Huge{j}}}                                      
(3.90, 0.0085) & 24 & 1.1317(3) & 1.0630(4) & 1.245(3)  &  0.188(1)&  1.54(2) &  -0.60(2) \\	            
{\phantom{\huge{l}}}\raisebox{-.2cm}{\phantom{\Huge{j}}}                                      
(3.90, 0.0100) & 24 & 1.1310(3) & 1.0632(4) & 1.240(5)  &  0.1850(8)&  1.48(3) &  -0.59(4) \\	            
{\phantom{\huge{l}}}\raisebox{-.2cm}{\phantom{\Huge{j}}}                                      
(3.90, 0.0030) & 32 & 1.1301(2) & 1.0615(3) & 1.234(3)  &  0.1820(7)&  1.54(4) &  -0.54(4) \\	            
{\phantom{\huge{l}}}\raisebox{-.2cm}{\phantom{\Huge{j}}}                                      
(3.90, 0.0040) & 32 & 1.1306(3) & 1.0621(3) & 1.238(6)  &  0.1841(7)&  1.52(3) &  -0.59(3) \\ \hline
{\phantom{\huge{l}}}\raisebox{-.2cm}{\phantom{\Huge{j}}}                            
(4.05, 0.0030) & 32 & 0.9411(2) & 1.0518(6) & 1.215(7)  &  0.147(1)&  1.69(3) &  -0.60(2) \\			
{\phantom{\huge{l}}}\raisebox{-.2cm}{\phantom{\Huge{j}}}                            
(4.05, 0.0060) & 32 & 0.9420(3) & 1.0534(5) & 1.240(10) &  0.146(2)&  1.60(4) &  -0.54(6) \\			
{\phantom{\huge{l}}}\raisebox{-.2cm}{\phantom{\Huge{j}}}                            
(4.05, 0.0080) & 32 & 0.9419(2) & 1.0519(4) & 1.218(9)  &  0.146(1)&  1.65(3) &  -0.57(5) \\ \hline	
{\phantom{\huge{l}}}\raisebox{-.2cm}{\phantom{\Huge{j}}}                            
(4.20, 0.0065) & 32 & 0.7807(3) & 1.0479(4) & 1.222(8)  &  0.116(1)&  1.79(3) &  -0.59(3) \\			
{\phantom{\huge{l}}}\raisebox{-.2cm}{\phantom{\Huge{j}}}                            
(4.20, 0.0020) & 48 & 0.7789(4)	& 1.0463(6) & 1.209(5)  &  0.114(2)&  1.72(6) &  -0.56(3) \\			
 \hline 
\end{tabular}}}
{\caption{\footnotesize  \label{tab:2} Detailed results for the hadronic quantities discussed in this paper, computed on each lattice data set specified in tab.~\ref{tab:1}. Note that $m_{\eta_c}$ and $f_{J/\psi}$ are given in lattice units. }}
\end{table}
These results should be now extrapolated to the physical limit ($m_{\rm sea}\equiv m_q\to 0$, $a\to 0$) by
\bea\label{extrap-V} 
V(0) = V(0)^{\rm cont.} \left[ 1 +  b_{V} m_{q} + c_{V} {a^2\over (0.086\ {\rm fm})^2}\right]\,.
\eea
from which we then obtain the two following values:
\bea\label{final-V} 
V(0)  = \left\{ \begin{array}{ll}
          1.937(34) &\quad \mbox{\rm (with\ $\beta=3.8$)}\,,\\
         &\\
          1.941(35)  &\quad \mbox{\rm (without\ $\beta=3.8$)}\,,\end{array} \right. 
\eea 
depending on whether or not we include the results obtained on our coarsest lattices in the continuum extrapolation. 
A few more comments are in order:
\begin{itemize}
\item The values of the $J/\psi\to \eta_c\gamma$ form factor computed on our lattices do not depend on the light sea quark mass, $b_{V}\simeq 0$. 
\item Although the smallness of discretization errors in tmQCD is guaranteed by construction [they are ${\cal O}(a^2)$], they are quite large for the form factor $V(0)$. More specifically, from the fit of our data to eq.~(\ref{extrap-V}) we get $c_V=-23(2)\%$. Note however that in this case the results obtained at $\beta=3.8$ do not modify the result of the continuum extrapolation, i.e. the formula~(\ref{extrap-V}) adequately describes all our results for $V(0)$. This can be appreciated from the plot provided in fig.~\ref{fig:6}.
\item The above results are obtained by using the semi-analytic ratio $R_{\rm sa}$. By repeating the same analysis with $R_{\rm num}$ we verify the same features concerning the ${\cal O}(a^2)$ effects and after extrapolating the results computed on all our lattices we obtain $V(0)=1.903(48)$, entirely consistent with the results quoted in eq.~(\ref{final-V}).
\end{itemize}
Our final result is:
\bea\label{FINALV}
V(0)=1.94(3)\left(^{+0}_{-4}\right)\,.
\eea
where the second error is a difference between the central values obtained by using the semi-analytic and numerical ratios discussed above.

We now turn to the discussion of the form factor $F_1(0)$, relevant to the $h_c\to \eta_c\gamma$ decay, as defined in eq.~(\ref{def-vectorFH}). To that end we compute the three point correlation functions 
\bea\label{Cijk}
C_{ijk} (\vec {q};t)&=&  \sum_{\vec x,\vec y} \langle  T_{ij}^\dag(0) J^{\rm em}_k(x)  P (y) \rangle \  e^{i\vec {q}( \vec x- \vec y)}  \nn\\
  &=&  - \langle \sum_{\vec x,\vec y}{\rm Tr}\left[  S_c^\prime(y;0) \gamma_i  \gamma_j  S_c (0,x)\gamma_k  S_c^{ \vec{\tilde \theta} }(x,y) \gamma_5 \right]\rangle \,.
\eea
Keeping in mind that we consider $h_c$ at rest and for $\vec {q}= -(1,1,1)\times \tilde \theta_0/L$, with $\tilde \theta_0$ tuned as indicated in eq.~(\ref{eq:angle2}), we can easily isolate the term proportional to $F_1(0)$ by combining:
\begin{align}
C_d(\vec {q};t)&={1\over 3}\left[  C_{123} (\vec {q};t) + C_{231} (\vec {q};t)+ C_{312} (\vec {q};t)\right]\,,\nn   \\ 
&\cr
 C_o(\vec {\tilde q};t)&={1\over 6}\left[  C_{131}  (\vec {q};t) + C_{212}  (\vec {q};t)+ C_{323} (\vec {q};t)\right.\nn\\
&\left. \qquad + C_{232}  (\vec {q};t) + C_{313} (\vec {q};t)+ C_{121} (\vec {q};t) \right] ,\nn\\
&\cr
C_{F_1} (\vec {q};t)&= C_d(\vec {q};t) - C_o(\vec {q};t) \nn\\
&\to \, {{\cal Z}_P^S  \over 2 E_{\eta_c}}e^{-E_{\eta_c} (T/2 - t)} \times  i \ m_{h_c} F_1(0)
\times  {  {\cal Z}_T^S\over 2 m_{h_c} }e^{- m_{h_c} t}
\end{align}
where, as before, in the last line we show the result of the spectral decomposition when all operators are sufficiently separated. By fitting 
\bea\label{eq:Rh}
R_{\rm h}(t) = {{\rm Im}[C_{F_1}(\vec {\tilde q};t)] \over {\cal Z}_P^S {\cal Z}_T^S} \times 4 m_{h_c}  E_{\eta_c}\  e^{m_{h_c} t + E_{\eta_c} (T/2 - t)}\,,
\eea
to a constant we obtain $m_{h_c} F_1(0)$. The results for the form factor $F_1(0)$, as obtained from all of our lattice data sets, are presented in tab.~\ref{tab:2}. These results too need to be extrapolated to the continuum and chiral limits, in a way similar to eq.~(\ref{extrap-V}). 
\begin{itemize}
\item The error on $F_1(0)$ computed on each of our lattices is under $10$\%, but is nevertheless twice larger than those we have when computing $V(0)$. This is expected as the signal for $h_c$ is much harder to tame. For the same reason we could not compute the form factor by employing the numerical method, i.e. by dividing by the two-point correlation functions. 
\item Contrary to the case of $V(0)$, the discretization effects on $F_1(0)$ are small (c.f. fig.~\ref{fig:6}). From the fit of our data to the form similar to eq.~(\ref{extrap-V}) we find $c_{F_1}\approx  2$\%. We get:
\bea\label{final-F1} 
F_1(0)  = \left\{ \begin{array}{ll}
          -0.57(2) &\quad \mbox{\rm (with\ $\beta=3.8$)}\,,\\ 
         &\\
          -0.57(3)  &\quad \mbox{\rm (without\ $\beta=3.8$)}\,,\end{array} \right. 
\eea 
\item Like in the case of $V(0)$, within the accuracy of our data, the form factor $F_1(0)$ is insensitive to the variation of the light sea quark mass.   
\item Our final result  is
\bea\label{FINALF1}
F_1(0) = -0.57(2)(1)\,,
\eea
where the second error is our estimate of the uncertainty due to the method for extracting the form factor from the correlation functions. In the case of $V(0)$ we were able to estimate that error from the difference between the central values obtained by using the semi-analytical~(\ref{eq:rsa}) and numerical ratios~(\ref{eq:rnum}). That difference turned out to be less than $2\%$. We add the same error to $F_1(0)$ leaving its sign free.
\end{itemize}

\subsection{$J/\psi$ annihilation constant}
In addition to the above quantities we also computed the annihilation constant $ f_{J/\psi}$ defined as,
\bea
\langle 0\vert \bar c (0) \gamma_\mu c(0)
\vert J/\psi (p,\lambda) \rangle = f_{J/\psi} m_{J/\psi} e_\mu^\lambda \,, 
\eea
which enters decisively in the expression for the well measured electronic width $\Gamma(J/\psi\to e^+e^-)$. Above, $e_\mu^\lambda$ stands for the polarization vector of $J/\psi$.   
$ f_{J/\psi}$ is computed along the same lines discussed in our previous paper~\cite{Becirevic:2012ti}, namely from the fit of the correlation function $C^{J/\psi}_{ii}(t)$ from eq.~(\ref{eq:0}) to the form
\bea
\label{r1}
C_{ii}^{J/\psi}(t) \xrightarrow[]{\displaystyle{ t\gg 0}} \; \frac{\cosh[  m_{J/\psi} (T/2-t)]}{ m_{J/\psi} }  \left| Z_A(g_0^2) \langle 0\vert \bar c(0) \gamma_i c(0)
\vert J/\psi  (\vec 0, \lambda) \rangle \right|^2 e^{- m_{J/\psi} T/2}
\;,
\eea
where the  non-perturbatively determined $Z_A(g^2_0)$ are those listed in tab.~\ref{tab:1}.~\footnote{ Note that in the three point functions we use $\bar \psi_c {\tau^{3}\over 2} \gamma_i \psi_c$, which is invariant under the same axial rotations that leave the twisted mass QCD action invariant  and is the invariant vector current multiplicatively renormalized by $Z_V(g_0^2)$. In the two point function, instead, $\bar \psi_c {\tau^{a}\over 2} \gamma_i \psi_c$ ($a=1,2$) is used which, at the maximal twist, corresponds to the axial current (in the twisted-mass basis) and is therefore renormalized by $Z_A(g_0^2)$~\cite{fr}.}
 In practice we of course combine the correlation 
functions in which both interpolating operators are smeared and those in which one operator is local and the other one is smeared. 
The results are given in tab.~\ref{tab:2}, which after extrapolating to the continuum by using a form analogous to eq.~(\ref{extrap-V}) lead to
\bea\label{Fpsi}
f_{J/\psi}=414\pm 8 ^{+9}_{-0}\ \mev\,.
\eea
The continuum extrapolation is smooth ($c_f=5(2)\%$) and the result is obtained by using all our lattices. If we leave out the results obtained at $\beta=3.8$, the resulting $f_{J/\psi}$ is larger by $9$~MeV, which is the second error quoted above. Note also that $f_{J/\psi}$ depends very mildly on the light sea quark mass ($b_f = 0.5(2)~\gev^{-1}$).  
\section{\label{sec-5}Phenomenology }
\subsection{Decays of $J/\psi$}
By using our result~(\ref{Fpsi}) we can compute the electronic width of $J/\psi$ as
\bea
\Gamma(J/\psi \to e^+e^-) = {4\pi\alpha_{\rm em}^2\over 3 m_{J/\psi} } {4\over 9} f_{J/\psi}^2 = 5.8(2)(1)~{\rm keV}\qquad [5.55(14)(2)~{\rm keV}]^{\rm exp.},
\eea
where we used $\alpha_{\rm em}=1/134$~\cite{alphaEM}, symmetrized the error bars, and quoted the experimentally established result~\cite{PDG} to better appreciate the agreement between our lattice result and experiment. 
Concerning the radiative decay $J/\psi \to \eta_c\gamma$, by inserting our value (\ref{FINALV}) in eq.~(\ref{widthPSI}) we get
\bea\label{eq:ourPSI}
\Gamma(J/\psi \to \eta_c\gamma) = 2.64(11)(3)~{\rm keV}\qquad [1.58(37)~{\rm keV}]^{\rm exp.},
\eea
where for the physical result we used the measured ${\rm Br}(J/\psi\to \eta_c\gamma)= (1.7\pm 0.4)\%$ and the full width $\Gamma_{J/\psi}= 92.9\pm 2.8$~keV~\cite{PDG}, as well as the physical values of $m_{J/\psi}=3096.92(1)$~MeV, and $\Delta=116.6\pm 1.2$~MeV. 
Had we used our $\Delta =112\pm 3$~MeV, the resulting $\Gamma(J/\psi \to \eta_c\gamma)$  would have been $10\%$ smaller.

Although somewhat lower than the quark model result, $\Gamma(J/\psi \to \eta_c\gamma) = 2.85$~keV~\cite{QMsummary}, our lattice result obviously gives a larger $J/\psi\to \eta_c\gamma$ decay rate, and the agreement with experiment is only at $2\sigma$. The effective theory approach of ref.~\cite{nora} and the QCD sum rule analyses~\cite{alex} succeeded at getting lower value for the decay rate of this decay, but with large uncertainties. Note that the dispersive (model independent) approach of ref.~\cite{shifman} predicted, $\Gamma(J/\psi \to \eta_c\gamma) = 2.2\div 3.2$~keV, many years ago. All these results agree with ours too, except that we have smaller and controlled uncertainties.  We hope more effort on the experimental side will be devoted to clarify the disagreement among various experiments, including the recent ones. For example, a study of this decay at BESIII  would give us a very valuable information. Recent result at KEDR suggested a larger value for the branching fraction ${\rm Br}(J/\psi\to \eta_c\gamma)=(2.34\pm 0.15\pm 0.40)$~\%~\cite{KEDR}, which would result in  $\Gamma(J/\psi \to \eta_c\gamma) = (2.2\pm 0.6)$~keV, in very good agreement with our result~(\ref{eq:ourPSI}).

As for the other lattice calculations of this form factor, we note that the quenched result of ref.~\cite{lattice-radiative-1}, $V(0)=1.85(4)$, is only slightly lower than ours, while the one obtained at single lattice spacing with $N_{\rm f}=2$ light flavors in ref.~\cite{lattice-radiative-2}, $V(0)=2.01(2)$,  is larger than our values at $\beta=4.05$ listed in tab.~\ref{tab:2}. Apart from different methodology, a notable difference is that  the authors of ref.~\cite{lattice-radiative-2} used the point-split electromagnetic current that does not require renormalization, whereas we use the local current on the lattice that is properly renormalized by non-perturbatively determined $Z_V(g_0^2)$.  Keep in mind that our final results is obtained after the extrapolation to the physical limit of the results obtained at several lattice spacings and for several different values of the light sea quark mass. 

\subsection{$h_c  \to \eta_c \gamma$}

$h_c$ escaped the experimental detection for a long time and only recently CLEO succeeded to isolate this state~\cite{Rosner:2005ry} and observed that its prominent mode is precisely $h_c\to \eta_c\gamma$, the branching fraction of which was later accurately measured at the BESIII experiment, with a result: ${\rm Br}(h_c\to \eta_c\gamma)= (53\pm 7)\%$~\cite{Ablikim:2010rc}. We obviously cannot compute the branching ratio on the lattice, but with our form factor result~(\ref{FINALF1}) we can compute the decay width using eq.~(\ref{widthH}). We get
\bea
\Gamma(h_c \to \eta_c\gamma) = 0.72(5)(2)~\mev\,.
\eea
This result can be combined with the measured ${\rm Br}(h_c\to \eta_c\gamma)$ to estimate the width of the $h_c$ state. We obtain:
\bea
\Gamma_{h_c}= {\Gamma(h_c \to \eta_c\gamma) \over {\rm Br}(h_c\to \eta_c\gamma)} = 1.37\pm 0.11\pm 0.18\ \mev\,,
\eea
where the first error comes from our determination of the form factor $F_1(0)$, and the second one reflects the experimental uncertainty in the branching ratio. Notice also that we symmetrized the error bars. This constitutes a prediction that would be interesting to check against the actual experimental measurement once the latter becomes available.~\footnote{  It will also be interesting to see the physics results of the effective theory developed in ref.~\cite{vairo}. }

To compare our estimate $F_1(0)=-0.57(2)$ with other lattice results we convert the value of the value reported in ref.~\cite{lattice-radiative-1} to our dimensionless form factor and obtain  $F_1(0)=-0.53(3)$, which agrees very well with our result.~\footnote{More specifically, the relation between our $F_1(0)$ and $\hat E_1(0)$, defined in ref.~\cite{lattice-radiative-1} is: $F_1(0)=a_t \hat E_1(0)/m_{h_c}$, with $a_t=6.05(1)$~GeV, and $a_t \hat E_1(0)=-0.306(14)$. }  Similar conversion of the result of ref.~\cite{lattice-radiative-2}  would result in $F_1(0)=-0.33(1)$, much smaller value than ours, whether we compare it with the values we obtain at $\beta=4.05$ or the one in the continuum limit. 
 
\section{\label{sec-6}Summary}

In this paper we presented results of our analysis of the radiative decays of charmonia by means of QCD simulations on the lattice. Using the twisted mass QCD with $N_{\rm f}=2$ dynamical flavors at several small lattice spacings we were able to smoothly extrapolate the relevant form factors to the continuum limit. 

\begin{itemize}
\item  We used the twisted boundary conditions to make sure that we extract the physical form factors, i.e. at $q^2=0$. We checked that at every lattice spacing explored in this paper our data indeed reproduce the latticized energy-momentum relation given in eq.~(\ref{eq:freeboson}). We also showed that the so-called {\sl kinetic} and {\sl rest} masses converge to the same value in the continuum limit, but that at fixed lattice spacing they both have discretization errors that are equal in size but different in sign.
\item We computed the hyperfine splitting and obtained
\bea
\Delta = m_{J/\psi}-m_{\eta_c} =112\pm 4\ \mev\,,
\eea 
and showed that it depends very mildly on the sea quark mass, with a slope being positive. From our computation of $R_{h_c}$ in eq.~(\ref{eq:resR}) we obtain $m_{h_c}=3.537(32)$~GeV, in good agreement with $m_{h_c}^{\rm exp.}=3.525$~GeV.
\item We computed the hadronic form factor relevant to the $J/\psi\to \eta_c\gamma$ (M1) transition and found
\bea
V(0) = 1.92(3)(2)\,,
\eea
which is larger than the one we would infer from the measured $\Gamma(J/\psi \to \eta_c\gamma)$, although compatible at the $2\sigma$ level. We found that the discretization effects are large and negative, but that they are adequately described by the linear function in $a^2$.
\item Our result for the $h_c \to \eta_c\gamma$ (E1) transition form factor is
\bea
F_1(0) = -0.57(2)(1)\,,
\eea
which mildly depends on the lattice spacing. Within the  uncertainties quoted  above, both our form factors are insensitive to the change of the sea quark mass. After combining our  $F_1(0)$ with the measured  ${\rm Br}(h_c\to \eta_c\gamma)$ we deduced the value of the width, $\Gamma_{h_c}=  1.37(22)$~MeV.
\item In addition to the above, we also computed the annihilation constant 
\bea 
f_{J/\psi}=418\pm 8 \pm 5\ \mev\,,
\eea
which agrees with the measured decay width $\Gamma(J/\psi \to e^+e^-)$. 
\end{itemize}
In the above results we do not make any estimate of the size of systematic uncertainty due to the omitted $s$ and $c$ quarks in the sea. In ref.~\cite{Li} it was claimed that the contributions from dynamical charm might be important. That point will be numerically assessed from the analysis similar  to the one presented in this paper but on the set of gauge field configurations that include $N_{\rm f}=2+1+1$ dynamical quark flavors. We also emphasize that our results are obtained without inclusion of the OZI suppressed contributions. Their impact appears to be small in $J/\psi\to e^+e^-$ decay, but their size in the radiative decays is unknown. They were  neglected and the associated uncertainty cannot be estimated without actually attempting to compute the corresponding disconnected diagrams on the lattice. Such a computation would be very welcome.

The  strategy employed in this paper can be applied to compute the much needed $F_1^{h_b\to \eta_b}(0)$ which we plan to do in the near future. 
\vspace{1.7 cm}

\section*{Acknowledgments}
We thank the members of the ETM Collaboration for discussions and for making their gauge field configurations publicly available, Vittorio Lubicz for comments on the manuscript, Emi Kou for discussion and for drawing our attention to ref.~\cite{Ablikim:2010rc},  Nora Brambilla, Christine Davies and Alexander Khodjamirian for valuable comments. Computations are performed using GENCI (CINES) Grant 
2012-056806.

\end{document}